# Effect of polarization fatigue on the Rayleigh coefficients of ferroelectric lead zirconate titanate thin films: experimental evidence and implications

X.J. Lou,[1][*] H.J. Zhang,[1] Z.D. Luo,[1] F.P. Zhang,[2] Y. Liu,[3] Q.D. Liu,[4] A.P. Fang,[5] B. Dkhil,[3] M. Zhang,[1] X.B. Ren,[1] H.L. He[2]

[1]*Multi-disciplinary Materials Research Center, Frontier Institute of Science and Technology, and State Key Laboratory for Mechanical Behavior of Materials, Xi'an Jiaotong University, Xi'an 710049, P. R. China*

[2]*National Key Laboratory of Shock Wave and Detonation Physics, Institute of Fluid Physics, CAEP, Mianyang 621900, P. R. China*

[3]*Laboratoire Structures, Propriétés et Modélisation des Solides, UMR 8580 CNRS-Ecole Centrale Paris, Grande Voie des Vignes, Châtenay-Malabry Cedex 92295, France*

[4]*State Key Laboratory for Strength and Vibration of Mechanical Structures, Xi'an Jiaotong University, Xi'an 710049, P. R. China*

[5]*Department of Physics, School of Science, Xi'an Jiaotong University, Xi'an 710049, P. R. China*

**Abstract:**

The effect of polarization fatigue on the Rayleigh coefficients of ferroelectric lead zirconate titanate (PZT) thin film was systematically investigated. It was found that electrical fatigue strongly affects the Rayleigh behaviour of the PZT film. Both the reversible and irreversible Rayleigh coefficients decrease with increasing the number of switching cycles. This phenomenon is attributed to the growth of an interfacial degraded layer between the electrode and the film during electrical cycling. The methodology used in this work could serve as an alternative *non-destructive* way for evaluating the fatigue endurance and degradation in dielectric properties of ferroelectric thin-film devices during applications.

---

[*] Email: xlou03@mail.xjtu.edu.cn or xlou03@163.com

Polarization fatigue in a ferroelectric (FE) material is defined as the progressive shrinkage of its hysteresis loop and the reduction in its switchable polarization during repetitive electrical cycling. The problem of electrical fatigue is one of the most serious failure issues for FE devices, which causes degradation of their ferroelectric (switching) and dielectric properties, and consequently hinders their full commercial applications.[1, 2] Although extensive experimental and theoretical investigations have been carried out to tackle this problem during the past decades, the origin of polarization fatigue remains elusive. The proposed phenomenological scenarios and microscopic explanations include, but by no means limited to, domain blocking by electronic charges,[3] generation of a low-dielectric-constant interfacial layer,[4] domain wall pinning by agglomerates of oxygen vacancies,[5, 6] domain nucleation inhibitions,[7] and the local phase decomposition caused by switching-induced charge injection (LPD-SICI model).[8, 9] However, the question as to which scenario/model plays a dominant role during fatigue is still controversial. Therefore, more investigations (especially the experimental ones) are required to reveal the physics behind this phenomenon.

On the other hand, a FE material is also a piezoelectric and high $K$ dielectric material. Therefore, apart from non-volatile memory applications, FE materials have also been widely used in other functional devices, such as integrated capacitors, energy storage devices, microwave devices, and so on.[10] For these applications, high dielectric constant of FE materials is required to remain stable upon repetitive bipolar/unipolar electrical cycling during use. Surprisingly, most of previous studies on electrical fatigue primarily concentrate on the degradation in the FE hysteresis-loop properties of FE devices with the number of switching cycles. And investigations of the change in their *dielectric* properties (e.g., the so-called reversible and irreversible Rayleigh parameters) during electrical cycling remain scarce. With the growth of interest in a variety of applications for high $K$ FE materials,[11] gaining a better understanding of the evolution of dielectric behavior during electrical bipolar/unipolar cycling (or during the charging/discharging process for high-energy storage devices) is highly desirable for predicting the performance and lifetime of such devices.

For a FE material, the measured dielectric response consists of both the intrinsic (reversible) and extrinsic (irreversible) contributions. The intrinsic (reversible) contribution originates from the lattice or ionic response from single FE domains, while the extrinsic (irreversible) contribution comes from polarization switching and domain wall (or phase boundary) movement process. Under "*intermediate field*" conditions, the field-dependent dielectric permittivity could be well described by the empirical Rayleigh law.[12-20] According to the Rayleigh law, the relationship between dielectric permittivity and electric field can be written as:

$$\varepsilon_r(E_0) = \varepsilon_r(0) + \alpha E_0, \qquad (1)$$

where $\varepsilon_r(E_0)$ is the measured dielectric permittivity under $E_0$, which is the amplitude of the electric field oscillation ($E_0 \ll E_c$, where $E_c$ is the coercive electric field), $\varepsilon_r(0)$ is the reversible Rayleigh coefficient (also called initial dielectric permittivity) due to the intrinsic lattice response and $\alpha$ is the irreversible Rayleigh coefficient. The contribution of $\alpha E_0$ to the observed dielectric response stems from irreversible domain wall (or phase boundary) motion, which usually takes place at relatively high fields. As a result, by monitoring the evolution of both the reversible and irreversible Rayleigh parameters with the number of switching cycles, we could obtain a better understanding and estimation of the changes in the intrinsic lattice response and domain wall mobility at different fatigue stages. Hopefully, the present work may provide further insights into the mechanisms of polarization fatigue in FEs.

Moreover, the methodology and technique used in this work may provide an alternative non-destructive approach for evaluating the fatigue properties of FE devices and the degradation in their dielectric properties upon repetitive bipolar/unipolar electrical cycling, the latter of which may be useful for assessing the lifetime and reliability of high $K$ FE materials during applications.

The $Pb(Zr_{0.3}Ti_{0.7})O_3$ (PZT) thin film of ~280 nm in thickness used in this work was deposited on $Pt(111)/Ti/SiO_2/Si(100)$ substrate by using a sol-gel spin coating technique. Details about the solution preparation and deposition conditions can be found elsewhere.[21, 22] The phase of the film was identified by X-ray diffraction (XRD, Shimadzu), which shows that the film is polycrystalline, and of pure perovskite structure without any secondary phases. The film thickness was measured by using a Field-Emission Scanning Electron Microscope (FE-SEM, Su-8010, Hitachi). Platinum top electrodes of $100 \times 100$ μm$^2$ in size were deposited on the film by *dc* sputtering via TEM grids. The polarization-electric field (*P-E*) loops were recorded with a triangle wave of 1 kHz and the fatigue properties of the film were measured with bipolar triangle wave of $10^5$ Hz by using a Radiant ferroelectric workstation. Dielectric permittivity tests were conducted using an Agilent E4980A *LCR* meter.

Fig 1 shows the general feature of the ferroelectric and dielectric properties of the PZT thin film. Fig 1(a) displays the hysteresis loops of the film recorded at various applied electric fields and 1 kHz in frequency. From Fig 1(a), one can see that our film exhibits excellent ferroelectric properties. In specific, the remanent polarization $P_r$ is *ca.* 27 μC/cm$^2$, the saturated polarization $P_s$ is *ca.* 54 μC/cm$^2$, and the coercive field $E_c$ is *ca.* 135 kV/cm under a high electric field of about 820 kV/cm. To detect the dielectric response of the film, the profile of dielectric permittivity versus the strength of electric field oscillation, recorded at different frequencies (i.e., 1 kHz, 10 kHz, 100 kHz), is illustrated in Fig 1(b). It can be seen that at all three frequencies, the dielectric permittivity initially remains constant and is almost independent on the strength of the electric field. When the field exceeds a *threshold* value, however, the dielectric permittivity increases linearly with increasing the amplitude of the applied electric field (that is, a Rayleigh-like behavior appears). As the field further increases, the permittivity shows higher order of nonlinearity [see Fig 1(b)]. According to the interpretations of the Rayleigh law in the literature, the dielectric nonlinearity observed in this work may be due to the extrinsic contribution coming from irreversible domain wall motions.[12-20] Meanwhile, it is found that the profile of dielectric permittivity decreases with increasing measurement frequency [see Fig 1(b)]. This phenomenon has been attributed to domain wall or interface pinning process, and could be modeled by a modified Rayleigh law, as proposed by Bassiri-Gharb and Trolier-McKinstry.[13]

To study the effect of polarization fatigue on the dielectric nonlinearity of the film, switching endurance measurements were conducted up to $9.01 \times 10^8$ cycles [see Figs 2(a) and 2(b)]. After a certain number of electrical cycles, dielectric nonlinearity studies were immediately carried out [Figs 3(a) and 3(b)]. Fig 2(a) shows the change in hysteresis loop of the PZT film with increasing fatigue number $N$. Apparently, one sees that the *P-E* loop is suppressed gradually as $N$ increases. To better appreciate the change in the remanent polarization during polarization fatigue, the curve of $2P_r$ versus $N$ is plotted in Fig 2(b). It is found that the remanent polarization $P_r$ decreases down to ~15% of its original value as the number of fatigue cycles $N$ increases up to $9.01\times10^8$ cycles. And the most significant loss in $P_r$ takes place at around $10^6$ cycles.

The change in the dielectric nonlinearity of the PZT film upon progressive electrical cycling is shown in Fig 3. Fig 3(a) shows the profiles of dielectric permittivity versus oscillating electric field of 100 kHz after 1, $3\times10^5$, $1.8\times10^6$, $1.14\times10^7$, $9.09\times10^7$, $9.01\times10^8$ switching cycles. In this work, the applied *ac* fields ranging from 0 to 54 kV/cm (well below $E_c\sim$ 135 kV/cm) were chosen to measure dielectric permittivity, as well as the Rayleigh-like behavior. From Fig 3(a), one can see that the dielectric permittivity curves of the film show two regimes, which are the 'low field' region and the 'Rayleigh' region, depending on whether the oscillating filed is higher than the threshold field (*ca*. 4 kV/cm) or not. In addition, it is found that the dielectric permittivity profile descends gradually with increasing the number of switching cycles. Note that the permittivity curve after $3\times10^5$ cycles shows a slight enhancement than the original one. This phenomenon may be attributed to the instability of the as-prepared film, and the subsequent stabilization and domain wall pinning/depinning process during the first $3\times10^5$ cycles.[23]

To further appreciate the evolution of dielectric nonlinearity with fatigue cycle number $N$, the curves of dielectric permittivity versus oscillating electric field shown in Fig 3(a) were fitted by using the Rayleigh relation shown in Eq (1). The fitted values of the initial dielectric permittivity $\varepsilon_r(0)$ and irreversible Rayleigh parameter $\alpha$ are shown in Table I. To better illustrate the change in both the reversible and irreversible Rayleigh parameters, $\varepsilon_r(0)$ and $\alpha$ as a function of cycle number $N$ are shown in Fig. 3 (b), respectively. From Table I and Fig. 3 (b), one sees that $\varepsilon_r(0)$ decreases considerably from its original value of 349.61 to 272.97 after $9.01 \times 10^8$ cycles, which corresponds to a reduction of ~22%. Meanwhile, the irreversible parameter $\alpha$ is reduced from 1.87 cm/kV to 0.82 cm/kV, demonstrating a more significant loss of ~56%.

It was reported that polarization fatigue in FEs might be due to pinning of domain walls by the redistribution of charged species (especially oxygen vacancies) during electrical cycling.[3, 5-7] This scenario could reasonably explain the loss of the irreversible Rayleigh parameter $\alpha$, which is the key parameter describing the change in domain wall mobility or the concentration of mobile domain walls during electrical fatigue.[12, 13, 19, 24, 25] However, this scenario faces difficulties in interpreting the reduction in $\varepsilon_r(0)$ after polarization fatigue in an explicit and natural manner, due to the lack of direct experimental evidence. Nevertheless, this effect may still contribute to the observed phenomena because one can imagine that point defects like oxygen vacancies may accumulate at the domain walls during polarization fatigue, and then they change the domain wall charges distribution, which can in turn modify the polarization within the domain and therefore $\varepsilon_r(0)$.

Additionally, among all the fatigue models and scenarios proposed in the literatures, no one explicitly invokes the relationship between the dielectric nonlinearity and polarization fatigue.[2] In this work, it is argued that the evolution of both the reversible and irreversible Rayleigh parameters may be *naturally* accounted for in the framework of the LPD-SICI model.[8, 9]

From the viewpoint of the LPD-SICI model, polarization fatigue is caused by local phase decomposition triggered by switching-induced charge injection at the film/electrode interface.[8,9] In this model, there is a critical parameter $E_{bc}J$, the power density generated by locally injected electrons, in which $E_{bc}$ is the depolarization field generated near the electrode by the bound charges at the tip of the needlelike domains during switching and $J$ is the injected current density (see Refs 8 and 9 for details). Under such condition, local phase decomposition may occur at the domain nucleation sites, which gives rise to the growth of an interfacial degraded layer with dielectric constant much lower than that of FE bulk. Therefore, for the fatigued sample, the whole system could be described by a in-series capacitors model,[26,27] $d/\varepsilon = d_i/\varepsilon_i + d_f/\varepsilon_f$, where $\varepsilon$, $\varepsilon_i$ and $\varepsilon_f$ are the dielectric permittivity of the whole capacitor, that of the degraded interface layer, and that of undamaged PZT FE film, respectively, while $d$, $d_i (\ll d)$, $d_f$, are the total film thickness, the thickness of the interface degraded layer, and that of the PZT film, respectively. Obviously, as the thickness of the degraded layer increases, the field seen by the FE PZT bulk is considerably reduced after fatigue, which leads to both a decrease in the reversible Rayleigh coefficient $\varepsilon_r(0)$ and a drop in the irreversible Rayleigh coefficient $\alpha$ [see Figs 3(a) and 3(b)]. Also, since local phase decomposition occurs only near the domain nucleation sites during polarization fatigue,[8,9] the decrease in the number of available domain nucleation sites, as well as the reduced switching probability of active FE domains due to the increase in the depolarization field,[28] also make polarization switching more difficult (and therefore further reduces the irreversible Rayleigh coefficient $\alpha$).

It is worth mentioning that the formation of a degraded interface layer in FEs during polarization fatigue has been known for a long time. For instance, Lou *et al* directly observed the formation of a pyrocholore-like layer underneath Pt electrode in PZT thin film by using micro-Raman spectroscopy.[8] Additionally, Chen and McIntyre found that Pb in the PZT film could react with Pt at the electrode/film interface, which leads to the growth of a non-FE interfacial layer during electrical fatigue.[29] Also, Nozaki *et al.* observed a transformation from Al electrode to $Al_2O_3$ under repetitive polarization switching in Al/vinylidene fluoride (VDF) oligomer FE capacitor, which has been attributed to the oxidation of Al electrode induced by the repeated charging and discharging process during polarization reversal.[30] Furthermore, the degradation and even partially melting of the electrode/film interface were directly observed by Balke *et al.* using SEM in PZT ceramic pellets after $3\times10^5$ switching cycles, and the authors attributed the suppression of hysteresis loop after fatigue to a field-screening mechanism.[31] All these observations, therefore, strongly support our arguments made above.

Finally, let us make a few remarks on the relationship between polarization fatigue and the evolution of the reversible/irreversible Rayleigh parameters and some important implications of this work. It is well known that electrical fatigue may cause changes in many properties of FE materials, e.g., the decrease in remanent/switchable polarization, change in domain pattern, variation in coercive field and charge transport properties, and so forth.[2] Conventionally, polarization fatigue in FEs is merely characterized by the reduction in remanent/switchable polarization. It is probably because that the change in coercive field $E_c$ is subtler than that in spontaneous polarization, that is, $E_c$ can be increased,[32] unaffected or slightly reduced after fatigue.[33, 34] Additionally, the change in leakage current with the number of switching cycles is unclear either. For instance, both the increase and decrease in carriers conductivity have been found in severely fatigued FE samples.[2] Therefore, neither coercive field nor leakage current can be used to characterize the fatigue properties of the sample.

In this work, a novel approach has been established to estimate the degree of polarization fatigue in FEs, that is, detecting the change in the reversible/irreversible Rayleigh coefficients with the number of fatigue cycles. Fig 4 shows the normalized $P_r$, $\varepsilon_r(0)$ and $\alpha$ as a function of the number of switching cycles. From Fig 4, one sees that all these three curves, though recorded at different field regimes, show a rather similar trend with increasing cycle number, with $P_r$ showing the most dramatic decrease, $\varepsilon_r(0)$ exhibiting the least dramatic decrease and $\alpha$ lying somewhere in the middle. Since the Rayleigh-law measurement is conducted under subswitching conditions, it causes *little* influence on the material and is therefore non-destructive. So, monitoring the Rayleigh coefficients with cycling number may provide another routine for polarization fatigue measurement.

Last but not least, as the interest in high $K$ FE materials grows, obtaining a better understanding of the change in dielectric properties during electrical cycling is indispensible for estimating the performance and lifetime of the devices. And the methodology used in this work may provide a non-destructive technique to evaluate the degradation of the dielectric behavior of FE devices, and therefore be useful for assessing the reliability and lifetime of high $K$ FE devices during operation.

In summary, we have studied the effect of polarization fatigue on the reversible/irreversible Rayleigh coefficients of the FE PZT film. It was found that both Rayleigh coefficients decrease as the number of switching cycles increases. We attributed this phenomenon to the growth of a degraded interface layer at the electrode/film interface. The methodology used in the present work may provide an alternative non-destructive approach for measuring polarization fatigue and the degradation in dielectric behavior of FE devices during applications.

This work was supported by the National Science Foundation of China (NSFC No. 51372195, 41372055 and 11272248), the Ministry of Science and Technology of China through a 973-Project (No. 2012CB619401), the Fundamental Research Funds for the Central Universities (2013JDGZ03), and National Key Laboratory of Shock Wave and Detonation Physics through a fund (No. LSD201201003). X.J. Lou would like to thank the "One Thousand Youth Talents" program for support. Y. Liu and B. Dkhil wish to thank the China Scholarship Council (CSC) for funding Y.L.'s stay in France.

**Figure captions:**

Figure 1 (a) Hysteresis loops of the PZT thin film, obtained under different *ac* electric fields. (b) The oscillating field dependence of dielectric permittivity measured at 1, 10 and 100 kHz.

Figure 2 (a) The change of hysteresis loop with increasing the number of switching cycles. (b) The doubled remanent polarization $2P_r$ as a function of cycle number.

Figure 3 (a) The oscillating field-dependent dielectric permittivity measured at 100 kHz under subswitching conditions ($|E| < 54$ kV/cm) recorded at different polarization fatigue stages, in which both the low-field region and the Rayleigh region can be clearly distinguished. (b) shows the evolution of the reversible and irreversible Rayleigh parameters [i.e. $\varepsilon_r(0)$ and $\alpha$] as a function of the number of fatigue cycles.

Fig 4. The normalized $P_r$, $\varepsilon_r(0)$ and $\alpha$ as a function of the number of fatigue cycles.

Table I. The reversible Rayleigh parameter $\varepsilon_r(0)$ and irreversible Rayleigh parameter $\alpha$ at different stages of polarization fatigue.

| No. of switching cycles ($N$) | $\varepsilon_r(0)$ | $\alpha$ (cm/kV) |
|---|---|---|
| 1 | 349.61 | 1.87 |
| $3 \times 10^5$ | 354.55 | 1.93 |
| $1.8 \times 10^6$ | 324.96 | 1.55 |
| $1.14 \times 10^7$ | 311.71 | 1.26 |
| $9.09 \times 10^7$ | 296.75 | 0.99 |
| $9.01 \times 10^8$ | 272.97 | 0.82 |

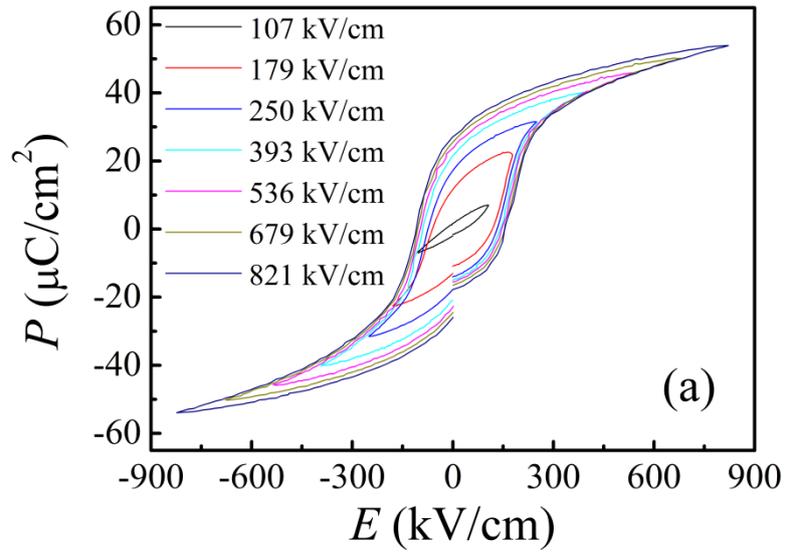

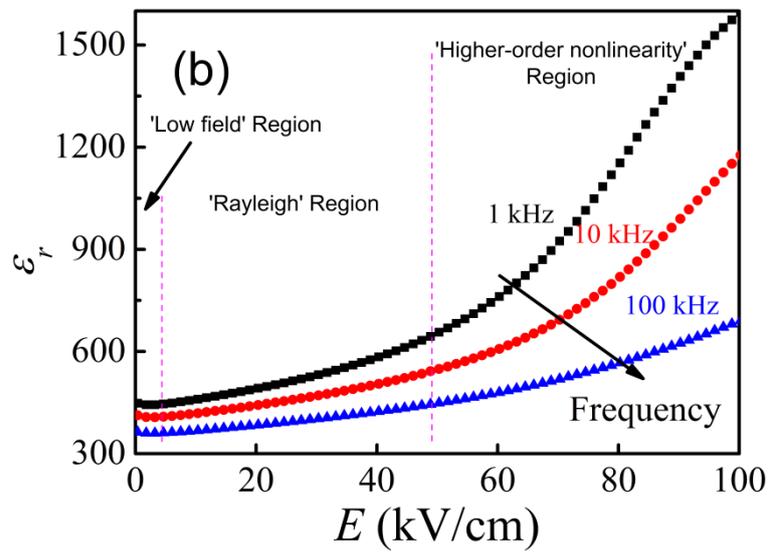

Figure 1

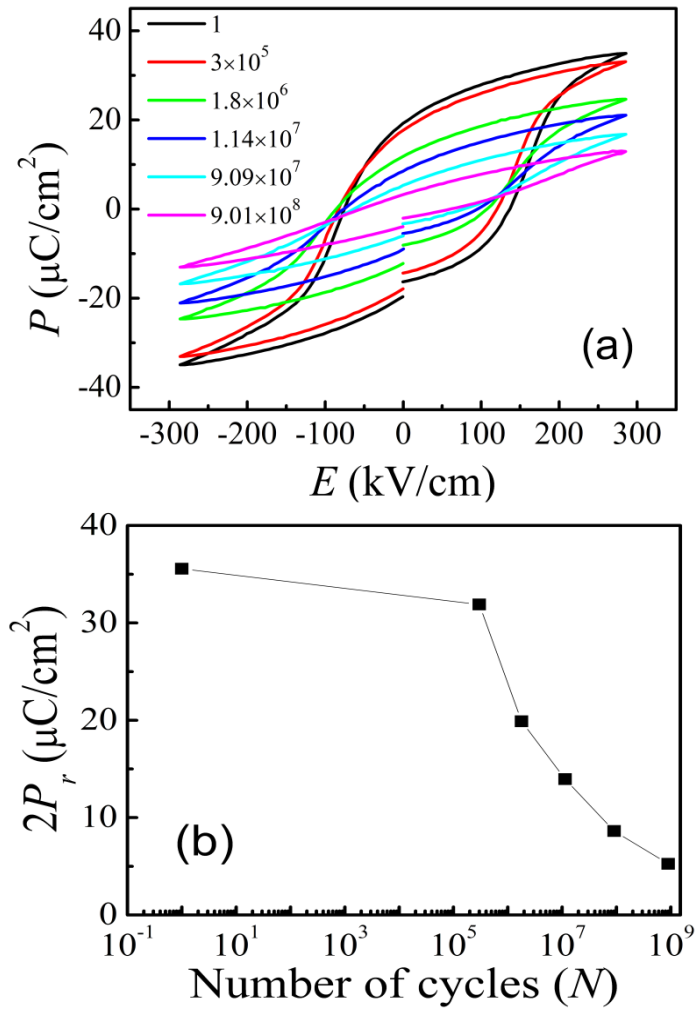

Figure 2

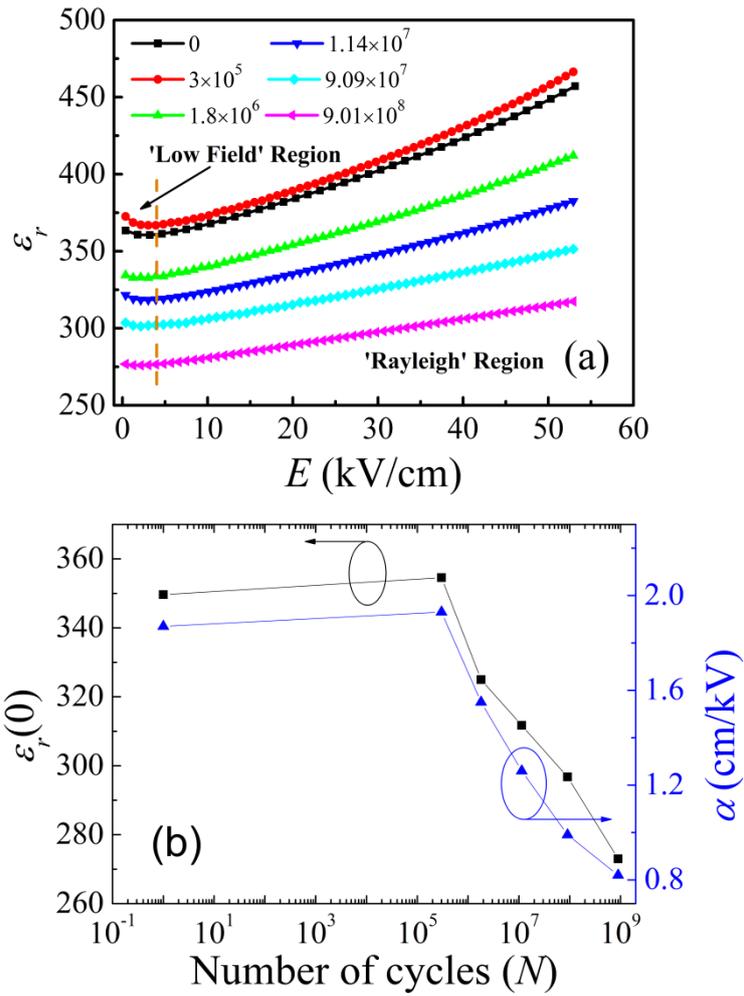

Figure 3

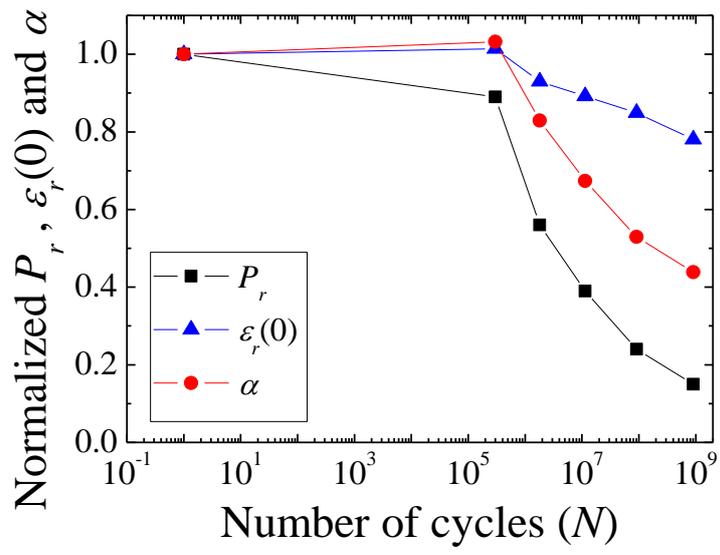

Fig 4